\documentclass[twocolumn,showpacs]{revtex4}  %Printversion
 \usepackage[latin1]{inputenc}
 \usepackage{amsmath}
 \usepackage{epsfig}
 \usepackage{subfigure}
 \usepackage{placeins}
 \usepackage{floatflt}
 \usepackage{amsmath}
 \usepackage{amssymb}
 \usepackage{upgreek}
 \usepackage{wrapfig}
 \usepackage[]{tocbibind}
 \usepackage{graphicx}

 \usepackage{ifthen}
 \usepackage{times}

\makeatletter
\newcommand\figcaption{\def\@captype{figure}\caption}
\makeatother

\newlength{\mylinewidth}
\begin{document}
%\linenumbers

%NO NEED TO TYPE FILE EXTENSIONS IN \includegraphics:
%\DeclareGraphicsExtensions{.jpg,.pdf,.peg,.mps,.eps,.ps}             %USE WHEN DOCUMENT IS PDFLATEX'ED!

\title{In-Medium Effects on K$^0$ Mesons in Relativistic Heavy-Ion Collisions}
\author{\footnotetext{corresponding authors:}  G.~Agakishiev$^{8}$, A.~Balanda$^{3,d}$, B.~Bannier$^{5}$, R.~Bassini$^{9}$,
D.~Belver$^{15}$, A.V.~Belyaev$^{6}$, A.~Blanco$^{2}$, M.~B\"{o}hmer$^{11}$, J.~L.~Boyard$^{13}$, P.~Cabanelas$^{15}$
% ...
}
\affiliation{
\footnotesize
\begin{center}
(HADES collaboration)\\
\end{center}
%\author{and\\
%C. Hartnack}
\\\mbox{$^{1}$Istituto Nazionale di Fisica Nucleare - Laboratori Nazionali del Sud, 95125~Catania, Italy}\\
\mbox{$^{2}$LIP-Laborat\'{o}rio de Instrumenta\c{c}\~{a}o e F\'{\i}sica Experimental de Part\'{\i}culas , 3004-516~Coimbra, Portugal}\\
\mbox{$^{3}$Smoluchowski Institute of Physics, Jagiellonian University of Cracow, 30-059~Krak\'{o}w, Poland}\\
\mbox{$^{4}$GSI Helmholtzzentrum f\"{u}r Schwerionenforschung GmbH, 64291~Darmstadt, Germany}\\
\mbox{$^{5}$Institut f\"{u}r Strahlenphysik, Forschungszentrum Dresden-Rossendorf, 01314~Dresden, Germany}\\
\mbox{$^{6}$Joint Institute of Nuclear Research, 141980~Dubna, Russia}\\
\mbox{$^{7}$Institut f\"{u}r Kernphysik, Johann Wolfgang Goethe-Universit\"{a}t, 60438 ~Frankfurt, Germany}\\
\mbox{$^{8}$II.Physikalisches Institut, Justus Liebig Universit\"{a}t Giessen, 35392~Giessen, Germany}\\
\mbox{$^{9}$Istituto Nazionale di Fisica Nucleare, Sezione di Milano, 20133~Milano, Italy}\\
\mbox{$^{10}$Institute for Nuclear Research, Russian Academy of Science, 117312~Moscow, Russia}\\
\mbox{$^{11}$Physik Department E12, Technische Universit\"{a}t M\"{u}nchen, 85748~M\"{u}nchen, Germany}\\
\mbox{$^{12}$Department of Physics, University of Cyprus, 1678~Nicosia, Cyprus}\\
\mbox{$^{13}$Institut de Physique Nucl\'{e}aire (UMR 8608), CNRS/IN2P3 - Universit\'{e} Paris Sud, F-91406~Orsay Cedex, France}\\
\mbox{$^{14}$Nuclear Physics Institute, Academy of Sciences of Czech Republic, 25068~Rez, Czech Republic}\\
\mbox{$^{15}$Departamento de F\'{\i}sica de Part\'{\i}culas, Univ. de Santiago de Compostela, 15706~Santiago de Compostela, Spain}\\
\mbox{$^{16}$Instituto de F\'{\i}sica Corpuscular, Universidad de Valencia-CSIC, 46971~Valencia, Spain}\\
\mbox{$^{17}$ Excellence Cluster Universe, Technische Universit\"{a}t M\"{u}nchen, Boltzmannstr.2, D-85748, Garching, Germany}\\
\\
\mbox{$^{a}$ also at ISEC Coimbra, ~Coimbra, Portugal}\\
\mbox{$^{b}$ also at Technische Universit\"{a}t Dresden, 01062~Dresden, Germany}\\
\mbox{$^{c}$ also at Dipartimento di Fisica, Universit\`{a} di Milano, 20133~Milano, Italy}\\
\mbox{$^{d}$ also at Panstwowa Wyzsza Szkola Zawodowa , 33-300~Nowy Sacz, Poland}}
\author{
\vspace{5mm}
C. Hartnack$^{18}$\\
\vspace{8mm}}
\affiliation{
\mbox{$^{18}$ Subatech, Ecole des Mines, 4 rue A. Kastler F-44307 Nantes, France.}\\
}
%\mbox{$^{f}$ also Extreme Matter Institute, GSI Helmholtzzentrum f\"{u}r Schwerionenforschung, D-64291 Germany}\\
\date{\today}
\begin{abstract}
We present the transverse momentum spectra and rapidity distributions of $\pi^{-}$ and K$^0_S$ in Ar+KCl reactions at a beam kinetic energy of 1.756 A GeV measured with the spectrometer HADES. 
The reconstructed K$^0_S$ sample is characterized by good event statistics for a wide range in momentum and rapidity.
We compare the experimental $\pi^{-}$ and K$^0_S$ distributions to predictions by the IQMD model.
The model calculations show that K$^0_S$ at low tranverse momenta constitute a particularly well suited tool to investigate 
the kaon in-medium potential. Our K$^0_S$ data suggest a strong repulsive 
in-medium K$^0$ potential of about 40 MeV strength.
 \end{abstract}
%accepted for publication in Phys. Rev. C %version of \today \\
\maketitle

\section{Introduction}

%T
Heavy ion collisions at relativistic energies  in  the SIS energy regime  (E =1-2 A GeV) allow to create rather dense nuclear systems up to few times the saturation density and this provides a favorable environment for the study of in-medium hadron properties. Within this context, expected medium effects on strange
particles have been in the focus of nuclear reaction studies at SIS energies for  the past two decades.
The predicted appearance of a kaon condensate in compressed nuclear matter \cite{Kap86} with its consequences for the understanding
of neutron star evolution \cite{Kol95} has emphasized the quest for a quantitative determination of the 
Kaon-Nucleon/Nucleus potential. So far, particular efforts have been devoted to the production of K$^{+}$ and K$^{-}$ mesons
and have provided phase space distributions, integral yields and angular distributions for a wide range of energies and collision systems.\\
The systematics of the experimental K$^+$ observables and in particular the results from sideward \cite{Cro00} and 
out-of-plane \cite{Shi98} flow analyses suggest a repulsive kaon-nucleus potential. 
Data from proton-induced reactions support a moderately repulsive potential for K$^+$  of the order of 20 MeV \cite{Sche06,Bes97} in agreement with theoretical calculations \cite{Bra97,Li96,Rud05,Fuc06}.\\
Results concerning the K$^-$ \cite{Sche06} are hampered by the low statistics available, and no settled quantitative conclusions could be drawn so far about the strength of the attractive potential.\\
Neutral kaons can shed additional light on the underlying question. They have the advantage that possible medium effects 
are not obscured by the Coulomb interaction. Similar to K$^+$, the K$^0$ in-medium potential is expected to be repulsive at these energies
as can be inferred from recent results extracted from pion-induced reactions \cite{Ben08}. 
Indeed, the comparison of the K$^0_S$ momentum distribution in the $\pi^-+C$ and $\pi^- +Pb$ reaction points to the existence of a repulsive KN potential of 20 $\pm$ 5 MeV at a normal nuclear density. \\
In this work we report on results for $\pi^-$ and $K^0_S$ extracted from Ar+KCl reactions at 1.756 A GeV. The high statistics data sample covers almost the entire phase space and allows a detailed analysis of the low-momentum component. 
For the first time, $p_t$ distributions for K$^0_S$ down to 50 MeV/c for the whole rapidity range could be measured. The study of the low-momentum region is well suited to access the K$^0$ potential in the nuclear medium since there repulsive effects are expected
to show up in a more pronounced way. Moreover, the spectral shape of the $p_t$ distribution allows quantitative conclusions, without requiring any absolute normalization necessary for descriptions which rely on measured yields only. 
The obtained $K^0_S$ data are compared to our results obtained for $K^+$ \cite{phiH09} and to theoretical calculations by the IQMD (Isospin Quantum Molecular Dynamics) transport model \cite{HaAi09} leading to an estimate of the strength of the repulsive K$^0$ in-medium potential.
\\
Our paper is organized as follows. Section II summarizes the experiment and section III addresses aspects of the particle identification method. 
This section presents also transverse mass spectra and rapidity distributions for $\pi^-$ mesons including a comparison to IQMD results. 
Section IV is devoted to the K$^0_S$ reconstruction procedure based on $\pi^- \pi^+$ pair decays. In section V we contrast the obtained K$^0_S$ 
distributions to K$^+$ data measured in the same reaction and compare them to results of IQMD simulations. In this section we discuss the
findings concerning the K$^0$ in-medium potential. We close with a summary in section VI.
\section{The Experiment}
The experiment was performed with the
{\bf H}igh {\bf A}cceptance {\bf D}i-{\bf E}lectron {\bf S}pectrometer (HADES)
at the heavy-ion synchrotron SIS
at GSI Helmholtzzentrum f\"ur Schwerionenforschung
in Darmstadt, Germany. A detailed description of the spectrometer is presented in
\cite{hadesSpectro}. \\
HADES consists of a 6-coil toroidal magnet centered on the beam axis and six identical detection sections located between the coils and covering
polar angles from $18^{\circ}$ to $85^{\circ}$. In the measurement presented here, the six sectors comprised
a gaseous Ring-Imaging Cherenkov (RICH) detector, four planes of
Multi-wire Drift Chambers (MDCs) for track reconstruction and two Time-of-Flight walls (TOF and TOFino), supplemented at forward polar angles
with Pre-Shower chambers. For each sector, the TOF and TOFino/Pre-Shower detectors are combined to a Multiplicity and Electron Trigger Array (META).
\\
A $^{40}_{18^+}$Ar beam of $\sim 10^6$ particles/s was incident on a four-fold segmented
KCl target with a total thickness corresponding to $3.3\%$
interaction length. A fast diamond start detector located upstream
of the target was used to determine the interaction time.
The data readout was started by a first-level trigger (LVL1) decision,
requiring an observed charged-particle multiplicity $MUL \ge 16$ in the TOF/TOFino detectors,
accepting approximately 35\% of the nuclear reaction cross section. This centrality selection translates into an average participants number $A_{part}$= 38.5$\pm2.7$, as described in \cite{phiH09,piH09}.
\section{Pion Identification}
\label{pions}
The reconstruction of K$^0_S \rightarrow \pi^- \pi^+$ decays requires a clean and unambiguous pion identification
within the high multiplicity track ensemble of each event.
Charged particles fire wires in the MDCs in front and behind the magnetic field and one or more hits in the META detector.  In the so-called cluster finder software \cite{PhD_Markert,hadesSpectro} track-segments are formed using the information from the two pairs of MDC planes. In the track-segment fitting procedures possible trajectories through the two  track segments and the  META hits are calculated.
For a detailed description of the tracking procedure see \cite{phiH09,PhD_Schmah}. After proper correlation of track segments and META hit points, 
particle momenta were calculated with a Runge Kutta integration of the trajectory in the magnetic field.
The graphical cuts utilized to select the $\pi ^{\pm}$ sample on the base of the $dE/dx$ versus momentum distributions are the same as shown in \cite{phiH09}. This selection allows to extract a high purity pion sample. 
Muons from pion decays ($\pi ^{\pm} \rightarrow \upmu^{\pm} +\upnu_{\upmu}, BR = 99,9\%, c\uptau = 7.8$ m) 
inside the HADES spectrometer constitute only a small fraction of the tracks and are mostly misidentified as pions
due to their small mass difference. Hence, the effective losses are in the order of few percent only, as shown in \cite{piH08}.
\\
The efficiency  of the MDC - $dE/dx$ cut and the purity of the selected sample have been extracted from experimental data, selecting particles fully reconstructed in the spectrometer and identified by the META as a reference and then verified using simulations, applying to the simulated tracks the same selection criteria as for the real data.
The average efficiencies, defined as the fraction of real pions surviving all the cuts are $\simeq 90\%$. The purity of this sample also corresponds to $\simeq 90\%$. While these values are rather independent of particle momentum
for $\pi ^-$ mesons, a significant reduction of the purity is observed for $\pi ^+$ with momenta p$ \ge 400$\,MeV/c due to contamination 
by misidentified protons. However, the $K^0_S$ reconstruction is only very weakly affected by the purity
of the high momentum $\pi^+$ sample.  
The acceptance of the HADES spectrometer and the efficiency
of the $\pi ^{\pm}$ reconstruction were determined by a full-scale GEANT3 \cite{geant} simulation.
A sample of pions generated with a flat distribution in rapidity ($-0.75<y_{c.m.}<0.75$) and transverse mass (0 $<m_t-m_\pi<550$ MeV/c$^2$) was propagated through the whole spectrometer yielding the geometrical acceptance from registered detector hits. 
The detection efficiency of the pion sample was obtained from the ratio of all emitted tracks inside the acceptance
to fully reconstructed tracks subjected to the same selection criteria as for the experimental data. 

 \section{Pion Results}
\begin{figure}[!t]
\begin{center}
\includegraphics[viewport= 55 13 536 720,angle=0,scale=0.41]{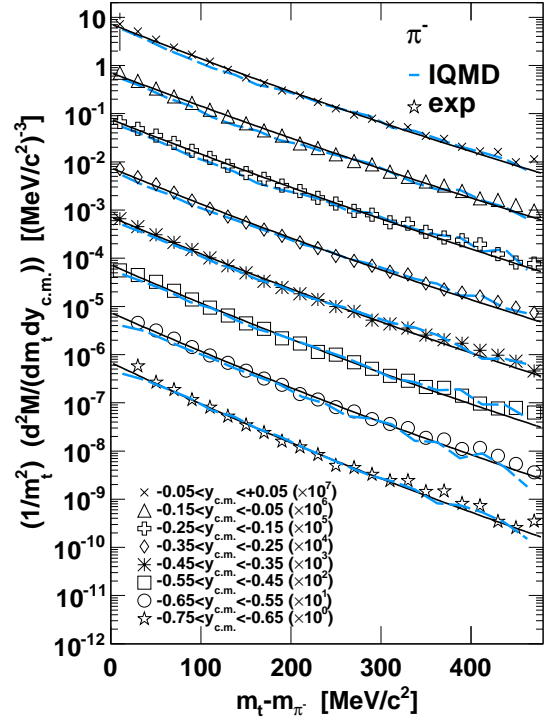}
\caption[]{(Color online). Reduced transverse mass distribution for different rapidity bins for the backward hemisphere for $\pi^-$ together with predictions by the IQMD model (dashed lines) and the fit (full lines) of the experimental data (symbols).}
\label{mtPiM}
\end{center}
\end{figure}
\begin{figure}[ht]
\begin{center}
\includegraphics[viewport= 55 13 536 400,angle=0,scale=0.40]{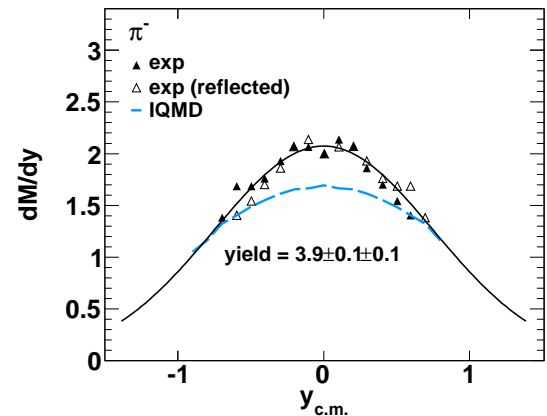}
\caption[]{(Color online). Rapidity distribution of $\pi ^-$ experimental data and reflected data points around mid-rapidity (full and empty triangles) together with the result from the IQMD calculation (dashed curve). The full curve shows a Gaussian fit of the experimental data.}
\label{rapPiM}
\end{center}
\end{figure}

Reduced transverse mass distributions for $\pi^-$ are shown in Fig.~\ref{mtPiM} for different center of mass rapidities 
(y$_{c.m.}$=y-y(c.m.), y(c.m.)= 0.86). The data are corrected for the efficiency and the spectrometer acceptance and are normalized
to the number of LVL1 triggers. The representation per transverse mass and rapidity unit divided by $m_t^2$
is chosen to ease the comparison with a Boltzmann distribution.
A fit according to
\begin{eqnarray}
\label{bolz_eqn}
\frac{1}{m_{t}^{2}} \frac{d^2M}{dm_{t}dy} &  = & C(y) \exp \left( -\frac{(m_t-m_0)c^2}{T_B(y)}\right) 
\end{eqnarray}
has been applied to the $m_t-m_{\pi}$ distribution in two adjacent  $m_t$ intervals. 
The pions mainly stem from the decay of the $\Delta(1232)$ resonance and are produced at different stages of the collision with varying 'hardness' as time proceeds. Hence, the measured $\pi$ spectrum is a superposition of different $\Delta$ generations.  
This explains why a single Boltzmann fit (\ref{bolz_eqn}) is not sufficient to match the data. It was found that a simultaneous two-slope Boltzmann fit with the transverse mass intervals chosen as $0 < m_t - m_{\pi} <$ 180 MeV/c$^2$ and 180 $< m_t - m_{\pi} < 500$ MeV/c$^2$, reproduces the data adequately.
The two $m_t$ ranges have been first fitted with two independent Boltzmann distributions yielding the start parameter values for the simultaneous fit. For the latter, the start values were allowed to vary up to a maximum of 10\%.
The resulting Boltzmann distributions are plotted in Fig.~\ref{mtPiM} as solid lines.
\\ 
As already shown in \cite{Rei07}, measured pion spectra could be reproduced reasonably well by IQMD calculations \cite{Bas95_0,Har98,Aic91,Bas93,Bas94_0,Bas94_1,Bas95_1}. The present HADES data are compared to IQMD calculations assuming a hard cut on the impact parameter $b<6$ fm which corresponds to a number of participants of 38.5$\pm$4.6.
The $\pi^-$ $m_t $ distributions obtained for the experimental data (symbols) are shown in Fig.~\ref{mtPiM} for different y$_{c.m.}$  bins together with the IQMD calculations (dashed lines) assuming the centrality selection $b<6$ fm.
\begin{figure}[!t]
\begin{center}
\includegraphics[viewport= 0 43 536 655,angle=0,scale=0.49]{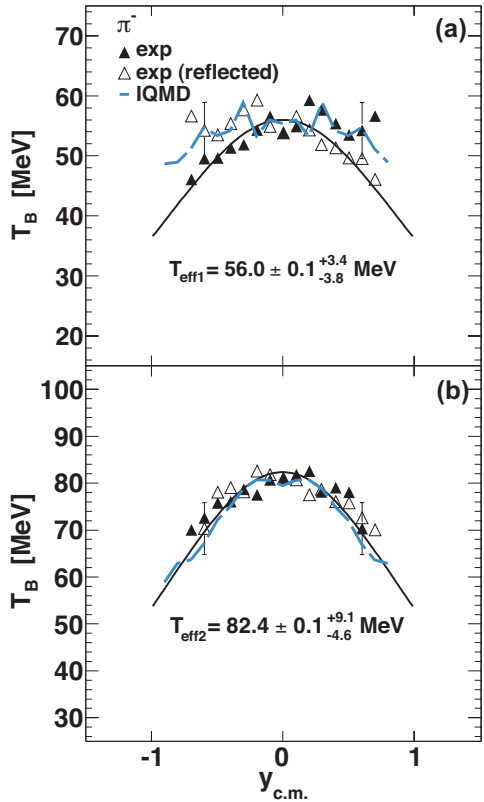}
\caption[]{(Color online). (Panel a)  Rapidity dependence of slope parameters obtained by fitting the $\pi^-$ $m_t$ distributions  from experimental  data (full symbols) and IQMD simulations (dashed curve) in the range  $0 < m_t - m_{\pi} < $ 180  MeV/c$^2$. (Panel b) Slope parameters obtained fitting the same distribution in the range 180 $ < m_t - m_{\pi} < $ 500 MeV/c$^2$. The values of the effective slopes $T_{\mathrm{eff}1/2}$ have been obtained employing function~(\ref{eqCos}).}
\label{TBPiM}
\end{center}
\end{figure}
The same fitting procedure has been applied to the IQMD data points. The obtained inverse slope parameters $T_{1,2}$ for experimental and simulated data are shown in Fig.~\ref{TBPiM} as a function of rapidity y$_{c.m.}$. 
We observe a systematic difference of 10-15\% between the slope factor $T_1$ determined in the forward and backward rapidity hemispheres which is not present in $T_2$. The agreement between data and model is of similar quality. \\
The full curves shown in Fig.~\ref{TBPiM} represent a fit to the experimental data according to 
\begin{equation}
T_B(y) = \frac{T_{\mathrm{eff}}} {\cosh(y-y_{c.m.})},
\label{eqCos}
\end{equation}
that allows to extract the effective slopes for the pion in the two m$_t$ ranges. The obtained values T$_{\mathrm{eff}1,2}$ are shown in the two panels of Fig.~\ref{TBPiM}.\\
Transverse momentum spectra of pions in Ar+KCl collisions 
at a very similar energy (1.808 A GeV)  have been published as kinetic energy spectra at 90$^{\circ}$ in the centre-of-mass frame \cite{San80}, which corresponds to mid-rapidity. The comparison with our data after proper normalization to the same number of participant nucleons shows agreement within the quoted errors.\\
The functions resulting from the fits to the pion transverse-mass distributions in the different rapidity bins in Fig.~\ref{mtPiM} are integrated over the interval $0 < m_t - m_\pi < \infty$ in order to obtain the yields as a function of rapidity. This integration method is justified by the large coverage of the  experimental $m_t - m_\pi$ distribution and the good agreement between the data and the fit function.
The resulting rapidity distribution of  ${\pi^-}$ is displayed in Fig.~\ref{rapPiM} together with the values obtained integrating the IQMD distributions. The two distributions are in agreement within 15\%; the largest deviation shows up in the mid-rapidity range.
The total yield of negatively charged pions was calculated from a Gaussian fit to the
$dN/dy$ distribution shown in Fig.~\ref{rapPiM}.  The systematic error on the pion
multiplicity has been estimated varying the boundaries 
used as start values for the transverse mass fits in the range from 120 to 300 MeV/c$^2$ and repeating for each step the same fitting procedure described above. We obtain M$_\mathbf{exp}(\pi^-)=3.9\pm0.1\pm0.1$ and M$_\mathbf{IQMD}(\pi^-)=3.7$ which are compatible within the error. 
Any systematic error derived from the comparison of the backward and forward distributions was found to be smaller than the statistical error.
These results have been confirmed by an independent analysis in \cite{piH09}. Taking as reference the pion multiplicities estimated in \cite{Rei07} for the Ar+KCl system at 1.756 GeV, agreement is achieved as well. \\
A comparison with earlier data and the IQMD model which is independent of the trigger details and impact parameter selection is possible by normalizing the pion yield to the number of participating nucleons. For the data presented here we find $\langle N_{\pi}\rangle/A_\mathbf{part} = 0.101\pm0.003\pm0.008$ (where the last error contains the uncertainty of the calculation of the A$_\mathbf{part}$value with UrQMD, $\approx$ 7\%). From IQMD we get $\langle N_{\pi}\rangle/A_\mathbf{part} = 0.105\pm 0.015$. From \cite{San80} for $\langle N_{\pi}\rangle$ = 5.6 and $A_\mathbf{part}$ =60  we obtain 0.093.
\section{K$^0_S$ Reconstruction}
\label{k0sRec}
\begin{figure}[!htb]
\begin{center}
\includegraphics[viewport= 35 13 536 440,angle=0,scale=0.40]{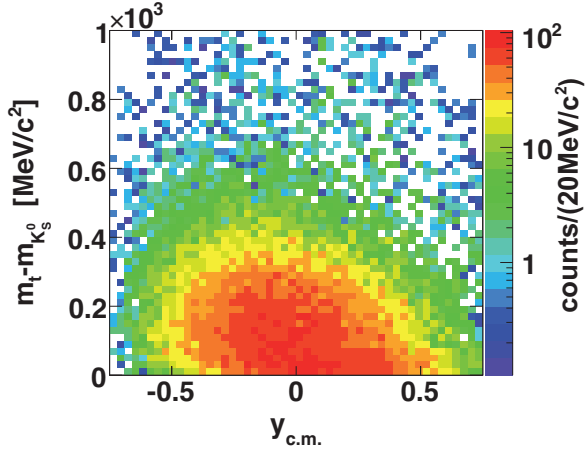}
\caption[]{(Color online). Distribution of reconstructed K${^0_S}$ yield as a function of the subtracted transverse mass $m_t-m_{K^0_S}$ and the center of mass rapidity y$_{c.m.}$. The color code refers to the number of kaons per bin.}
\label{k0sPS}
\end{center}
\end{figure}

\begin{figure}[!htb]
\begin{center}
\includegraphics[viewport= -5 13 536 430,angle=0,scale=0.645]{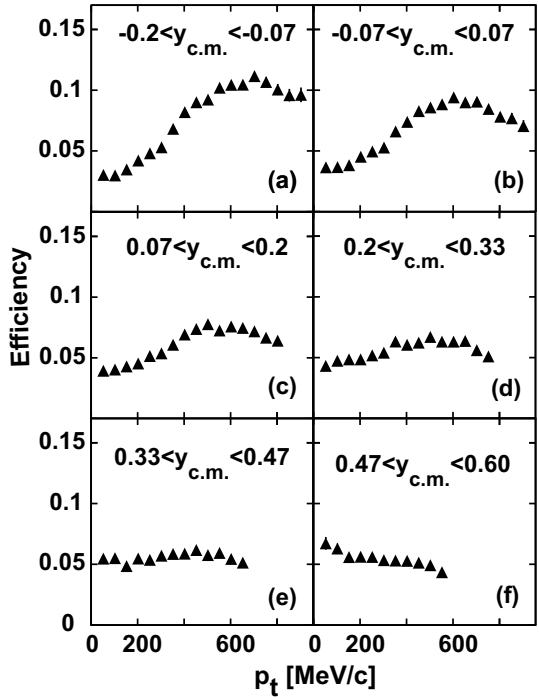}
\caption[]{K$^0_S$ reconstruction efficiency as a function of the transverse momentum $p_t$ and for different rapidity bins. The geometrical acceptance is not taken into account.}
\label{k0Eff}
\end{center}
\end{figure}

The K$^0_S$ meson (mean decay length $c\uptau$= 2.7 cm) decays into a $\pi ^+ + \pi^-$ pair with a branching ratio of 69\%. 
After selection of events with two identified pion tracks,
the invariant mass was calculated for each $\pi ^+-\pi^-$ combination.
In order to reduce the combinatorial background from uncorrelated $\pi ^+-\pi^-$ pairs and hence to
enhance the purity of the $K^0_S$ signal, various cuts on characteristic geometrical distances 
have been applied:
(1) the minimum distance between the two pion tracks ($d_{\pi^+-\pi^-}<$ 10 mm), 
(2) the distance of closest approach to the primary vertex for the two pion tracks ($d_0(\pi^+,\pi^-)\geq$ 6 mm and $d_0(K^0_S)<$ 10 mm),
and (3) the distance between the primary reaction and secondary decay vertex ($d(K^0_S-V)\geq$ 30 mm).
The method is discussed in detail in ref. \cite{PhD_Schmah}. 
\begin{figure}[t]
\begin{center}
\includegraphics[viewport= -35 33 436 640,angle=0,scale=0.44]{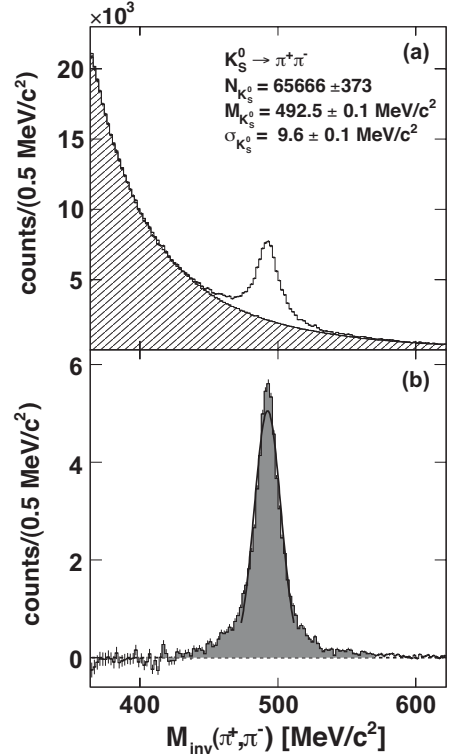}
\caption[]{Invariant-mass distribution of ${\pi^+}-{\pi^-}$ pairs (panel a).
The combinatorial background (shaded area) is obtained by the mixed-event technique.
The background-subtracted distribution (panel b) shows clearly the K$_S^0$ meson signal
(grey area with a Gaussian fit).}
\label{k0Mass}
\end{center}
\end{figure}
Fig.~\ref{k0Mass} (top) shows the resulting invariant-mass distribution of all $\pi ^+-\pi^-$ pairs. The peak corresponding to the K$^0_S$ signal is clearly visible on the top of the background which is reproduced using the mixed-event technique (dashed area in the top of Fig.~\ref{k0Mass}).  For the background analysis, only those events were chosen
for which the individual pion tracks originate from the same target segment of the four-fold KCl target stack.
 \\ The resulting K$_S^0$ signal after background subtraction is shown in Fig.~\ref{k0Mass} (bottom) and it can be fitted by the sum of two Gaussian distributions. A rather sharp distribution that contains most of the yield and a broader distribution that corresponds to the case where at least one of the two pion has undergone multiple scattering.
The Gaussian fit in Fig.~\ref{k0Mass} shows the result for the sharper distribution and corresponds to the following mean value and dispersion for the reconstructed mass:  $\langle m_{K^0_S}\rangle= 492.5 \, {\rm MeV/c}^2\,$ and $ \langle \upsigma_{K^0_S}\rangle= 9.3\, {\rm MeV/c}^2$. The width of the reconstructed signal and the signal-to-background ratio depend on the rapidity bin and vary in the range  $7\, {\rm MeV/c}^2<\upsigma_{K^0_S}< 12.4\, {\rm MeV/c}^2$ and $0.3<S/B<2.0$, respectively.
A total of about 65.700 K$_S^0$ mesons are identified in an interval of $\pm 3 \upsigma$ around the fitted mass peak. 
\\
The phase space distribution of the measured K$_S^0$ is shown in Fig.~\ref{k0sPS} as a function of the reduced transverse mass $m_t-m_{K^0_S}$ and center of mass rapidity y$_{c.m.}$.
From the plot it is evident that the low transverse momentum region is covered with significant statistics for the whole rapidity range.\\
For the quantitative analysis, acceptance and efficiency corrections have been applied to the 
reconstructed K$^0_S$ signal in a similar way as described for the single $\pi^\pm$ tracks in section \ref{pions}.
A K$^0_S$ event generator with a flat distribution in rapidity ($-0.75<y_{c.m.}<0.75$) and transverse mass ($0<m_t - m_{K^0_S}<$900 MeV/c$^2$) has been used as the input of a full-scale simulation in order to evaluate the geometrical acceptance and the reconstruction efficiency.
The average geometrical acceptance for K$^0_S$  amounts to 20-25\% \cite{PhD_Schmah}, the reconstruction efficiency to $\simeq 5-10\%$. The latter
is shown in detail in Fig.~\ref{k0Eff} as a function of the transverse momentum $p_t$ for different rapidity bins. In addition, the branching ratio of the decay K$^0_S \rightarrow \pi^+ + \pi^- $ was corrected for.
\section{K$^0_S$ Results}
\subsection{Transverse Mass Spectra} 
\begin{figure}[!htb]
\begin{center}
\includegraphics[viewport= 55 33 536 720,angle=0,scale=0.40]{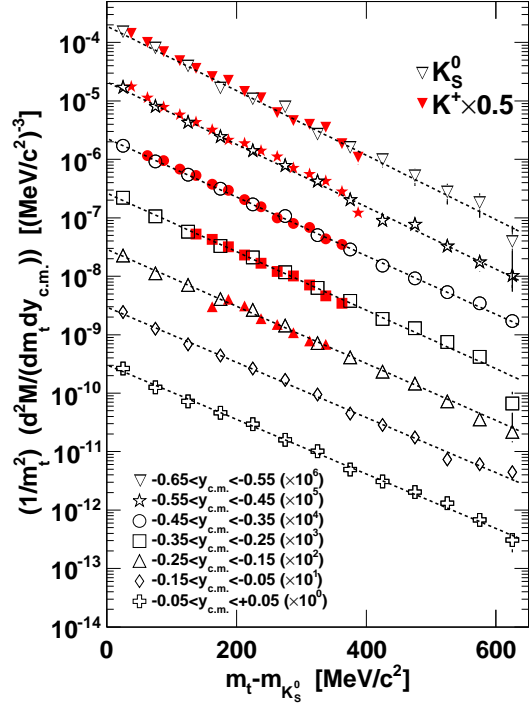}
\caption[]{(Color online). Transverse mass spectrum for different center of mass rapidity bins, corresponding to the backward hemisphere only, for K$^0_S$ (empty symbols), K$^+$ multiplied by a factor 0.5 (full symbols) and fits assuming a Boltzmann  parameterization (dashed lines).}
\label{k0Mt}
\end{center}
\end{figure}

\begin{figure}[!htb] 
\begin{center}
\includegraphics[viewport= 40 20 536 420,angle=0,scale=0.40]{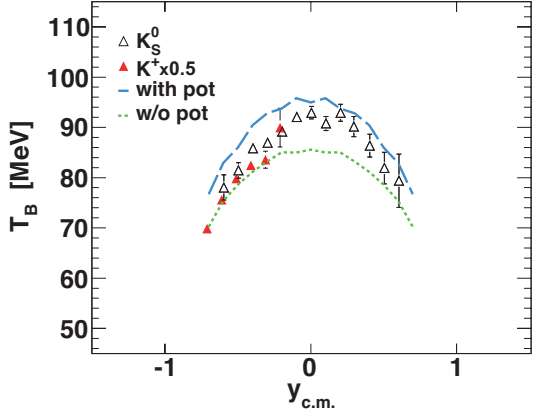}
\caption[]{(Color online). Rapidity dependence of slope parameters obtained by fitting the K$^0_S$ and K$^+$ $m_t$ distributions  from experimental  data (empty and full symbols respectively). The results by IQMD simulations employing a repulsive potential (dashed curve) and no potential (dotted curve) are shown as well. The errors are statistical only.}
\label{tempK0K+}
\end{center}
\end{figure}

The K$_S^0$ transverse mass spectra obtained after background subtraction and correction for acceptance and efficiency are plotted
in Fig.~\ref{k0Mt} (open symbols) for various y$_{c.m.}$ bins together with the K$^+$ spectra (full symbols).
 Since in isospin symmetric heavy-ion reactions the yields of K$^+$ and K$^0$ should be the same and since the K$^-$ yield is negligible compared to the K$^+$ yield we have $K^+ + K^-=K^0+\bar{K}^0= K_S^0+ K^0_L=2\cdot K_S^0$. For this reason, the data for the K$^+$ mesons (full symbols) in Fig.~\ref{k0Mt} have been multiplied by a factor 0.5. The quantitative comparison of these experimental spectra, that are found to be in good agreement, is a valuable cross-check of the analysis.
Neglecting final state Coulomb interactions, the dynamics of the interaction between K$^+$ and K$^0$
mesons and the nuclear medium should result in similar kinematic distributions. \\
Using the Boltzmann parameterization (\ref{bolz_eqn}), an inverse slope parameter $T_B(y)$  can  be determined as a function of the rapidity. The dashed lines in Fig.~\ref{k0Mt} shows the Boltzmann fit-functions. The resulting $T_B(y)$ values as a function of the rapidity obtained from the fit of the K$^0_S$ data are shown in Fig. \ref{tempK0K+} together with the results obtained for the K$^+$ \cite{phiH09} and by the IQMD simulations. The two curves in Fig. \ref{tempK0K+} refer to two different scenarios: 1) no in-medium potential (dotted lines in Fig.~\ref{rapK0K+}) and 2) a repulsive potential of 40 MeV (dashed lines). 
 Fitting  the $T_B(y)$ distribution with the function (\ref{eqCos}), the parameter $T_{\mathrm{eff}}$ can be extracted. It represents the inverse slope at mid-rapidity and corresponds to an effective temperature at the kinetic freeze-out stage. 

 The averaged $T_{\mathrm{eff}}$ for K$^0_S$ is found to be $92\pm2\,$MeV which agrees within the errors with the findings for K$^+$ in the same data set \cite{phiH09}.  
\subsection{Rapidity Distribution}
The fitted K$^0_S$ invariant transverse mass distributions are integrated within the whole interval $0 < m_t - m_{K^0_S} < \infty$ in order to obtain the meson yield per rapidity unit. To evaluate the precision of this integration, the integral of the fitting function has been compared to the integrated data points in the range $0 < m_t - m_{K^0_S} <$ 600 MeV/c$^2 $. 
We find an agreement between 0.5 and 2.5\% for all rapidity bins, except for the interval $-0.65<y_{c.m.}<-0.55$ for which the difference amounts 6.5\%.\\
\begin{figure}[!htb] 
\begin{center}
\includegraphics[viewport= 10 20 436 720,angle=0,scale=0.47]{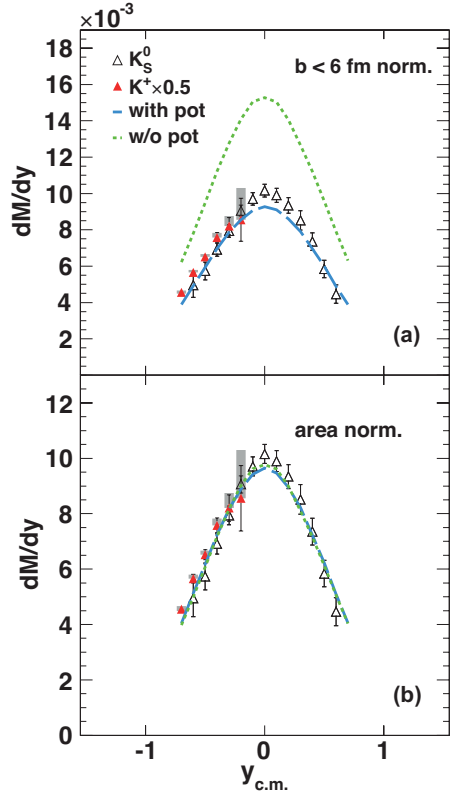}
\caption[]{(Color online). Rapidity distribution of K$^0_S$ (empty triangles) and K$^+$ mesons (full triangles from \cite{phiH09}) together with the distributions obtained with the IQMD calculation assuming a repulsive (dashed line) and no (dotted line) potential for two different normalization procedures (panel a and b, see text for details). The vertical boxes show the systematic errors associated to the K$^+$ data.}
\label{rapK0K+}
\end{center}
\end{figure}
The resulting rapidity distribution is shown in Fig.~\ref{rapK0K+} (empty symbols), together with the respective K$^+$ yields divided by a factor two (full symbols).  
The systematic error of the K$^{0}_{S}$ yield has been estimated by varying the geometrical cuts and the track quality selection in several combinations. The systematic errors shown for the K$^+$ distribution have been calculated as described in \cite{phiH09}.
Both experimental distributions overlap within the error bars.
Appropriate integration yields a total K$^{0}_{S}$ multiplicity of  $(1.15 \pm 0.05 \pm 0.09) \cdot 10^{-2}$ as compared to a total K$^+$ multiplicity of (2.8$\pm$0.2$\pm$0.1$\pm$0.1$)\cdot10^{-2}$ \cite{phiH09}, where the first and second (third) error are the statistical and systematic respectively. The differences in the tails of the rapidity distributions shown in Fig.~\ref{rapK0K+} lead to slightly different $4\pi$ yields.
Nevertheless the multiplicities, extracted with two independent analyses, agree within the error, confirming the quality of the K$^0_S$ data.
\\
Together with experimental distributions the IQMD results for the same two scenarios presented in Fig. \ref{tempK0K+} are shown.
In the top panel of Fig.~\ref{rapK0K+} the IQMD calculations ($b< 6$ fm) without any normalization are shown, while in the bottom panel the two curves are shown after a normalization to the total area of the experimental distribution. The IQMD calculation used here corresponds to the standard setting with $\alpha=$ 1.0 (see section \ref{compiqmd}) already shown in \cite{Fuc06,Foe07}.

Although the standard parametrization used in the IQMD calculations above 
has been successfully used for comparison to KaoS and FOPI data and thus
supports the results of this comparison, a direct conclusion on the optical 
potential from absolute yields is premature. It was shown in \cite{Har02}
that transport models using opposite assumptions on the optical potential but 
different parametrizations of poorly known production cross sections were able 
to reproduce the same rapidity distribution of kaons in central Ni+Ni events 
at 1.93 AGeV measured by FOPI and KaoS.

The calculations with and without potential have indeed very similar shapes (bottom panel of Fig.~\ref{rapK0K+}). The rapidity distribution contains the integrated information about  $p_t$ and assuming that the effect of the potential is momentum dependent, the differential study of the $p_t$ distribution can add important information. The study of the $p_t$ distributions between 50 and 800 MeV/c constitutes the key clue
of our approach to determine the in-medium $K^{0}$ potential, as described in the following section. \\
\subsection{Comparison with IQMD}
\label{compiqmd}
As shown in section \ref{pions}, the IQMD calculations are in reasonably good agreement with the pion spectra measured for the Ar+KCl reaction at 1.756 A GeV as far as the slopes are concerned, but differ up to 15\% in the absolute yield. These findings impose a caveat on the absolute normalization of the IQMD simulations on the experimental data. Starting with the same centrality selection used to obtain the pion spectra ($b<6fm$), a set of calculations with the IQMD model \cite{HaAi09} has been carried out, employing a repulsive K$^0$-nucleus potential of varying strength.  
\\
\begin{figure}[htb]
\begin{minipage}[b]{75mm}
      \begin{picture}(700,150)(0,0)
      \put(-18,-160.6){ \includegraphics[width=0.70\textwidth,height=1.49\textwidth]
          {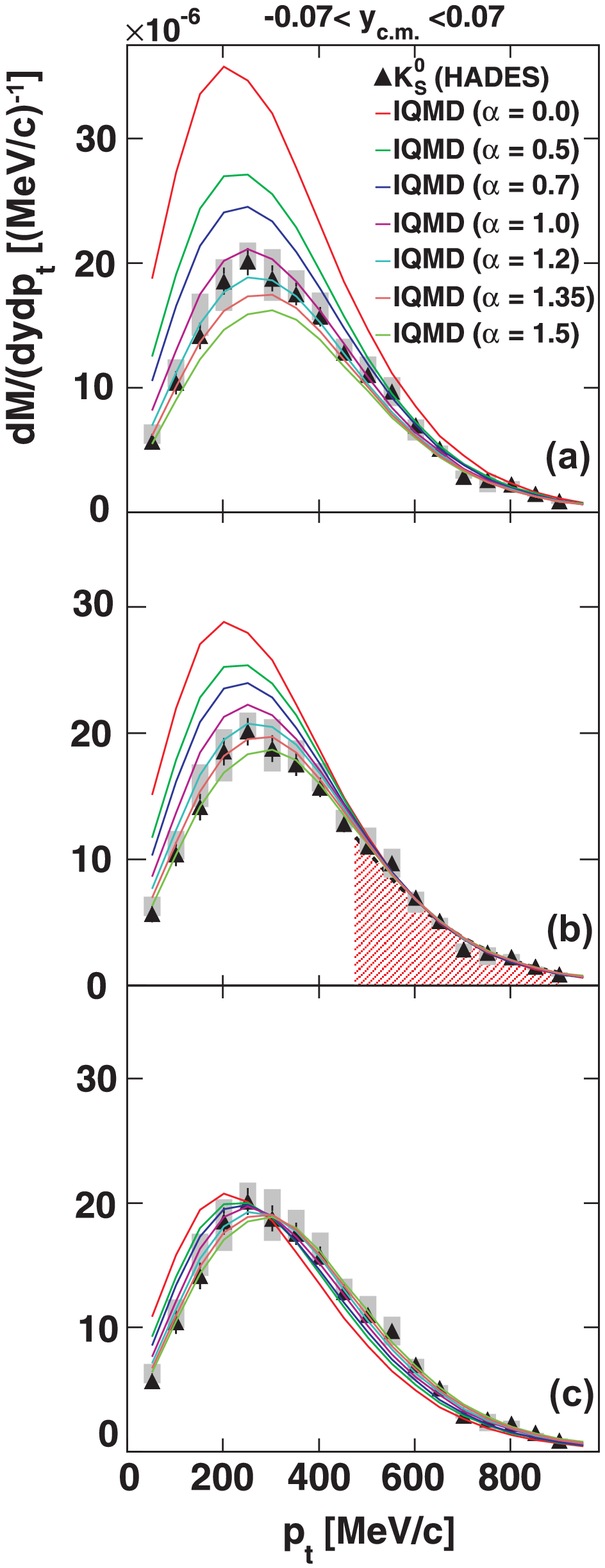}} 
          \end{picture}
\end{minipage}
 \begin{minipage}[t]{75mm}
      \begin{picture}(2,160)(0,0)
      \put(-7,0.6){ \includegraphics[width=.57\textwidth,height=1.49\textwidth]
          {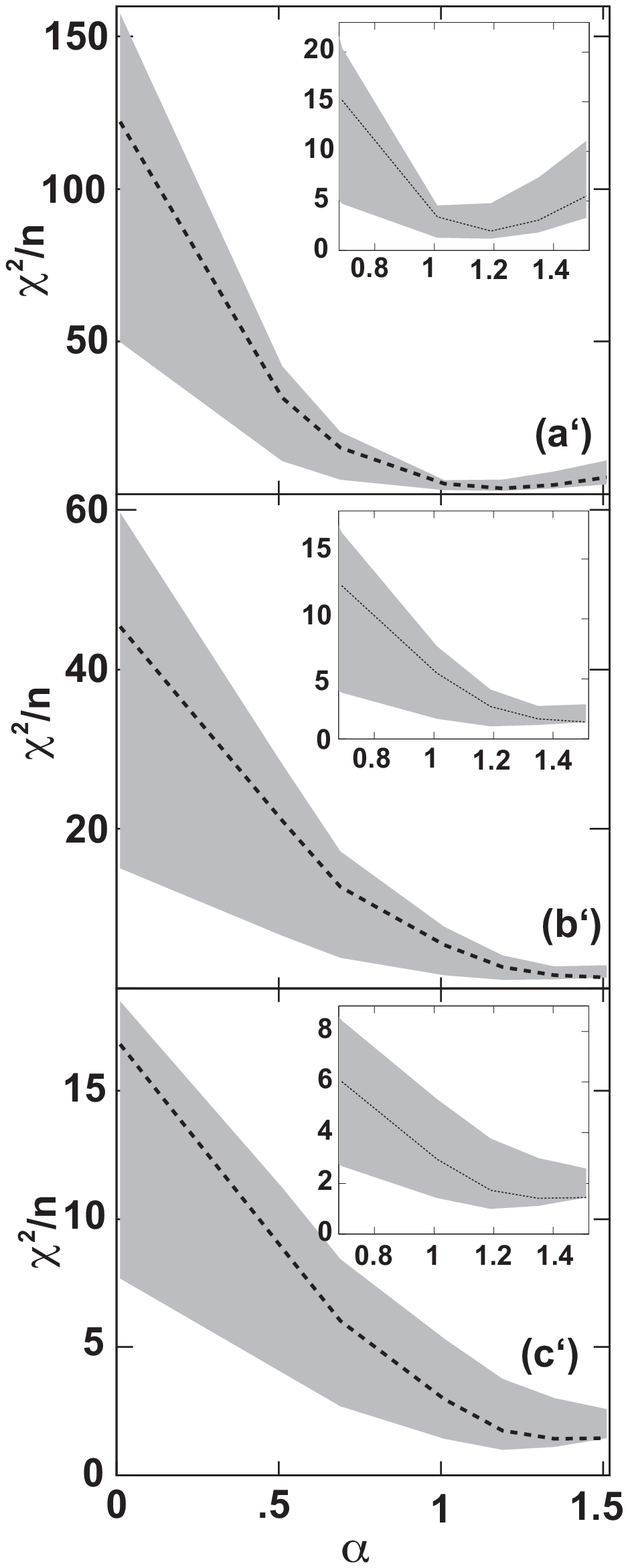}}
      \end{picture}
\end{minipage}   
   \caption{Color online. Left panels: $p_t$ distributions of K$^0_S$ at mid-rapidity (full symbols) compared with different calculations by the IQMD model. The different solid curves correspond to a variation of the parameter $\alpha$, which couples directly to the potential strength. The comparison is shown for three different normalization procedures (see text for details). Right panels: Normalized $\upchi ^2$ distribution as a function of the parameter  $\alpha$ extracted from the comparison of the IQMD calculations to the experimental data for three different normalization procedures (see text for details). The insets show a zoom around the minimum of the distribution.}
   \label{Kialpha}
    \end{figure}
According to \cite{Kor05}, K$^+ = \bar{s}u$ is a "good" quasi-particle with narrow width; the same is expected for K$^0 = \bar{s}d$. Its dispersion relation may be written as  $\upomega^2 = m_K^2 + k^2 + \Pi$ with $\upomega(k)$ as energy (momentum) and $\Pi$ as real part of the self-energy.
The latter one accounts for the influence of the ambient medium, thus leading to an effective in-medium mass $m* = \upomega(k=0)$.
In agreement with the low-density theorem, the K$^0$ effective mass should increase in nuclear matter, as for the K$^+$.
Another way to state this is to say that the kaons interact with surrounding nucleons via a Schr\"odinger-type potential.
An increase of m* is thereby related to a repulsive total potential. The easiest parameterisation is the linear Ansatz $m* = m_\rho + U(\alpha) \rho/\rho_0$ where $\rho (\rho_0)$ is the nuclear matter (saturation) density. For the IQMD settings \cite{HaAi09}, the parameter $\alpha$ is related to the potential U via U$(\alpha) \simeq U_0 + U'\alpha$ with $U_0~\approx$ 0.8 MeV and $U' \approx 38$ MeV \cite{HaAi09}. 
\\
In order to evaluate more quantitatively  the comparison between IQMD and experimental data, we have first focused on the mid-rapidity $p_t$ distribution and compared systematically the experimental data with different IQMD calculations obtained varying the parameter $\alpha$.
The left panels of Fig.~\ref{Kialpha} show the  K$^0_S$ p$_t$ distribution at mid-rapidity, together with the results of the IQMD calculations for different values of the parameter $\alpha$. The three panels correspond to three different ways of normalization of the IQMD calculations to the experimental data. Panel (a) in Fig.~\ref{Kialpha} shows the case where the IQMD curves for $b<6$ fm have not been normalized to the data, panel (b) shows the case where all the IQMD curves have been normalized to the high-$p_t$ tail ($p_t>475$ MeV/c) of the experimental distribution and panel (c) shows the case where the area underlying each curve has been normalized to the integral of the experimental distribution.
The error shown for the experimental data contains also the systematic contribution. This contribution has been evaluated varying the cuts for the K$^0_S$ selection for 14 different combinations.
A $\upchi^2$ analysis has been carried out applying a best fit of the different IQMD curves to the experimental data. 
The result corresponding to the three normalization procedures are shown in panels (a'), (b') and (c') of Fig.~\ref{Kialpha} where the error bands include both the statistical and systematic contributions. The inlets in the three panels show a zoom on the minimum region.
The minimum of the $\upchi^2$ distribution that corresponds to an optimal set of cuts for the K$^0_S$ is obtained for the following $\alpha$ values respectively: $1.13-0.12$, $1.37-0.2$ and $1.34-0.17$. 
\\
The asymmetric error on $\alpha$ has been evaluated taking the minimum $\upchi^2$ value +1 for each of the 14 different $\upchi^2$ distributions and reading the corresponding $\alpha$ value on the left side of the minimum. The maximal deviation of this value from the minimum of the dashed curves has been assigned to the asymmetric error. 
 Since IQMD calculations corresponding to $\alpha$ values higher than 1.5 lead to unphysical results, it is not possible to calculate the upper value of the $\alpha$ parameter. \\ 
One can see that even considering the three normalization methods simultaneously the lowest limit for $\alpha$ is 1, which correspond to a minimum repulsive potential of 38.7 MeV.\\
It has also to be mentioned that the IQMD calculations quoted in \cite{Foe07}, which interpret the whole K$^+$ KaoS systematics in favor of a soft EOS but do not draw any conclusions on the potential, used $\alpha=1.0$ as well. \\
This rather high value of the extracted potential is larger that the results reported in \cite{Ben08,Cro00,Rud05,Sche06,Nek02}, which reported a value of $20 \pm 5$ MeV. 
However, the potential values extracted so far from K$_S^0$ data are derived from $\pi^-$ +A reactions \cite{Ben08}. In this case the pion absorption happens on the nucleus surface, so that many of the produced kaons do not travel through the nucleus. Furthermore, the HADES data deliver higher statistics and accuracy in the measurement of the low p$_t$ range, which is more sensitive to the potential effects. Also, the results extracted from K$^+$ data in proton-induced reactions \cite{Nek02} test subnormal nuclear density and can not be directly compared with the data from heavy-ion collisions where average densities of 1.5-2 $\rho_0$ are reached.
In overall the nuclear environment resulting in the Ar+KCl reaction could lead to a stronger repulsive potential for K$^0$. \\
On the other hand, as extensively discussed in \cite{Fuc06,Kol05}, the different transport codes that have been employed to interpret the available data on kaons differ substantially in their implementation, in their elementary production cross-sections and in their treatment of the parameters connected to the strength of the potential. 
To get a consistent picture, the HADES K$^0_S$ data can be described by other models and the interpretation of the K$^+$ data previously measured in heavy-ion collisions \cite{Cro00} might be revised using updated versions of theoretical models and elementary cross-sections \cite{Kol05}. \\
\begin{figure}[!htb]
\begin{center}
\includegraphics[viewport= 10 13 536 720,angle=0,scale=0.54]{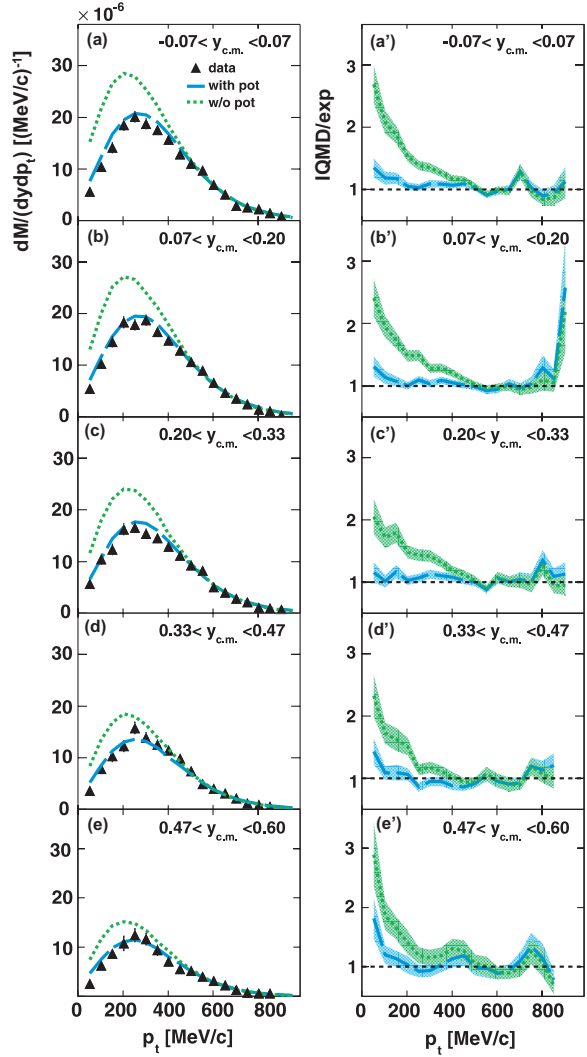}
\caption[]{Color online. Left panels: $p_t$ distribution of the experimental K$^0_S$ data (full triangles) together with the yields calculated by the IQMD model including a repulsive K$^0$-nucleus potential of 46.1 MeV (dashed curves) and without potential (dotted curves) for different rapidity bins. Right panels: ratio between the calculation by the IQMD model and the experimental data as a function of $p_t$ for different rapidity bins.}
\label{ExpIqmd}
\end{center}
\end{figure}
Taking as a reference the IQMD simulation corresponding to $\alpha$=0 and $\alpha$=1.2,  we have compared  them  to the experimental $p_t$ distributions 
as shown in Fig.~\ref{ExpIqmd}. 
The left panels of  Fig.~\ref{ExpIqmd} show the K$^0_S$ p$_t$ distributions for different rapidity bins, 
together with the IQMD calculations assuming either no repulsive potential (dotted curves) or a potential of about $46.1$ MeV ($\alpha=$ 1.2, dashed curves). 
 For this comparison the simulations are normalized to the high-momentum tail of the experimental distribution starting from $p_t=475$ MeV/c, since in this region the effect of the repulsive potential should be negligible. The displayed errors are only statistical.
 On the right panels of Fig.~\ref{ExpIqmd}, the ratio of the simulated to experimental yield is displayed for the two cases with and without a repulsive potential. The error band displayed for the ratios are again only statistical.
 \\
It is very interesting to notice that the IQMD simulation assuming a repulsive potential agrees rather well with the experimental data for all the rapidity bins. The calculations without the potential, on the other hand, overestimate the experimental data, especially in the low transverse momentum regime ($ <$ 400 MeV/c), as is clearly visible in the ratio plots. This behavior is slightly more evident at mid-rapidity but persists with the same trend for the whole rapidity range.
\\
We note that the $p_t$ spectra exhibit the most sensitive dependence on the K$^0$-nucleus potential. 
The rapidity distribution is sensitive to variations of the quantity $\alpha$ as well, but this effect is less pronounced and the 
conclusions drawn on the potential strength are strongly dependent upon the normalization of the simulated to experimental data. Hence the accurate study of the $p_t$ spectra delivers a new more powerful tool to extract quantitative information.

\section{Summary}
Rapidity and transverse momentum spectra of $\pi^-$ and K$^0_S$ mesons produced  in Ar+KCl collisions at 1.756 A GeV and
measured with HADES have been investigated. For the first time in this energy regime, K$^0_S$ spectra have been measured 
with high statistics and precision in almost the full phase space and, in particular, down to low momenta ($p_{min}\approx$ 50 MeV/c).
\\
The $\pi^-$ data have been shown and compared quantitatively to IQMD calculations, 
using an absolute normalization based on the selection of the impact parameter $b<6$ fm in the simulations and a centrality 
selection on the experimental data corresponding to the most central 35\% of the total cross section. 
The $\pi ^-$  rapidity density distribution shows a quantitative agreement with the IQMD model within 15\%.
\\
For K$^0_S$ mesons the transverse mass and the rapidity density distribution are compared to previously published K$^+$ data and agree
well with them, both in shape and yield.
The comparison of the K$^0_S$ rapidity distribution with IQMD calculations does not seem to be solid enough to extract reliable information about the potential, since the normalization is not certain and the calculations corresponding to different values of the potential deliver curves with the same shape.
The K$^0_S$ p$_t$ distributions have been found to be a better observable and  have been compared to calculations by the IQMD model assuming different strengths of the K$^0$-nuclear medium potential.
This comparison supports the existence of a rather strong repulsive potential of about 40 MeV.
These data are now available for further studies via the available transport models.
\subsection*{Acknowledgements}
% as for C+C pion paper
We gratefully acknowledge the useful discussions with J. Aichelin and H. Oeschler.\\
The HADES collaboration gratefully
acknowledges the support by BMBF grants 06TM970I, 06GI146I, 06FY171,
and 06DR135 (Germany), by GSI (TM-FR1, GI/ME3, OF/STR), by Excellence
Cluster of Universe (Germany), by grants GA
AS CR IAA100480803 and MSMT LC 07050 (Czech Republic), by grant KBN
5P03B 140 20 (Poland), by INFN (Italy), by CNRS/IN2P3 (France), by
grants MCYT FPA2000-2041-C02-02 and XUGA PGID T02PXIC20605PN
(Spain), by grant UCY-10.3.11.12 (Cyp\-rus), by INTAS grant
06-1000012-8861 and EU contract RII3-CT-2004-506078.

\end{document}